\documentclass[11pt,twocolumn]{article}
\usepackage[preprint]{acl}
\usepackage[T1]{fontenc}
\usepackage{amsmath,amssymb}
\usepackage{booktabs}
\usepackage{tabularx}
\usepackage{enumitem}
\usepackage{xurl}
\usepackage{microtype}

\title{Lightweight Query Routing for Adaptive RAG:\\
A Baseline Study on RAGRouter-Bench}
\author{
  Prakhar Bansal \\
  \texttt{prakharb13@gmail.com} 
  \And
  Shivangi Agarwal \\
  \texttt{shivangia@iiitd.ac.in} \\
}
\date{}

\begin{document}
\maketitle

\begin{abstract}
Retrieval-Augmented Generation pipelines span a wide range of
retrieval strategies that differ substantially in token cost and
capability. Selecting the right strategy per query is a practical efficiency
problem, yet no routing classifiers have been trained on
RAGRouter-Bench \citep{wang2026ragrouterbench},
a recently released benchmark of $7,727$ queries spanning four
knowledge domains, each annotated with one of three canonical
query types: factual, reasoning, and summarization.
We present the first systematic evaluation of lightweight
classifier-based routing on this benchmark.
Five classical classifiers are evaluated under three feature
regimes, namely, TF-IDF, MiniLM sentence embeddings
\citep{reimers2019sbert}, and hand-crafted structural features, yielding 15 classifier feature combinations.
Our best configuration, TF-IDF with an SVM, achieves a
macro-averaged F1 of $\mathbf{0.928}$ and an accuracy of
$\mathbf{93.2\%}$, while simulating $\mathbf{28.1\%}$ token savings
relative to always using the most expensive paradigm.
Lexical TF-IDF features outperform semantic sentence embeddings by $3.1$ macro-F1 points, suggesting that surface keyword patterns
are strong predictors of query-type complexity.
Domain-level analysis reveals that medical queries are hardest
to route and legal queries most tractable.
These results establish a reproducible query-side baseline
and highlight the gap that corpus-aware routing must close.
\end{abstract}

\section{Introduction}
\label{sec:intro}

RAG has become the dominant approach for grounding LLM outputs in
external knowledge \citep{lewis2020rag}, but the design space spans
multiple retrieval paradigms with very different cost profiles.
A simple dense-retrieval step (\textsc{NaiveRAG}) is fast and cheap;
an iterative pipeline (\textsc{IterativeRAG}) that alternates
retrieval and generation can handle complex queries but costs
3.5$\times$ more in tokens \citep{wang2026ragrouterbench}.
Most deployed systems apply one paradigm uniformly to every query.

Routing each query to the cheapest sufficient paradigm has clear
precedent.
\citet{jeong2024adaptiverag} trained a classifier to predict question
complexity and route among three retrieval strategies, showing that
lightweight routing can match always-expensive baselines.
The broader LLM routing literature shows similar results when
routing among model sizes \citep{ong2025routellm, hu2024routerbench,
ding2024hybridllm, chen2023frugalgpt}.

\citet{wang2026ragrouterbench} recently released RAGRouter-Bench,
the first benchmark designed specifically for RAG routing research.
It provides $7,727$ queries across four domains, each annotated with
one of three canonical query types (factual, reasoning,
summarization), alongside evaluations of five RAG paradigms.
Importantly, the benchmark establishes that paradigm applicability
is shaped by \emph{query-corpus interactions}, not query type alone.
To our knowledge as of April $2026$, no classifier has been trained on
this benchmark.
We fill that gap, establishing the first query-side classification
baselines and quantifying how much of the routing signal is
captured by query text features before corpus-side signals are
incorporated.

Our contributions are: (i) $15$ classifier--feature baselines on
RAGRouter-Bench using query text alone; (ii) a cross-domain
breakdown identifying where lightweight routing is most and least
reliable; and (iii) a post-hoc cost simulation showing that the
best high-accuracy configuration achieves $28.1\%$ token savings,
with an analysis of how cost and accuracy trade off across
feature regimes.

\section{Background}
\label{sec:background}

\subsection{RAGRouter-Bench}

RAGRouter-Bench \citep{wang2026ragrouterbench} comprises four
corpora: MuSiQue (Wikipedia, $3,356$ queries), QuALITY (literature,
$1,200$ queries), UltraDomain (legal, $1,277$ queries), and
GraphRAG-Bench (medical, $1,896$ queries).
The benchmark annotates each query with one of three canonical
query types: \emph{factual} (single-step fact lookup),
\emph{reasoning} (multi-hop cross-document inference), and
\emph{summarization} (corpus-level aggregation), encoded in the
dataset as \texttt{single\_hop}, \texttt{multi\_hop}, and
\texttt{summary} respectively.
The label distribution is imbalanced: factual queries account
for $52.9\%$ of all queries, summarization for $30.0\%$, and
reasoning for $17.1\%$.
Five RAG paradigms are evaluated with reported relative token costs:
\textsc{LLM-Only} ($1.0\times$), \textsc{NaiveRAG} ($1.4\times$),
\textsc{GraphRAG} ($2.1\times$), \textsc{HybridRAG} ($2.8\times$),
and \textsc{IterativeRAG} ($3.5\times$).

The paper's central finding is that no single paradigm universally
dominates: optimal paradigm selection is driven by
\emph{query-corpus interactions}, with both query type and
structural or semantic corpus properties shaping which paradigm
performs best \citep{wang2026ragrouterbench}.
This motivates our work as a query-side baseline study---
establishing what is achievable from query text alone, before
corpus-side signals are incorporated.

\subsection{Related Work}

\paragraph{Adaptive retrieval:} Adaptive-RAG \citep{jeong2024adaptiverag} trains a T5-Large
classifier on automatically derived complexity labels to route
among three retrieval strategies, and demonstrates that a
three-class query-complexity router can match always-expensive
baselines with substantially lower cost.
Our work extends this approach to RAGRouter-Bench with lighter
classifiers, a richer domain spread, and a comparison of feature
types.
Self-RAG \citep{asai2024selfrag} embeds retrieval decisions in
the generation process via reflection tokens; FLARE
\citep{jiang2023flare} uses generation probabilities as a
retrieval trigger; Probing-RAG \citep{baek2025probingrag} reads
LLM hidden states to make binary retrieve-or-not decisions.
SKR \citep{wang2023skr} routes based on the model's apparent
self-knowledge; CRAG \citep{yan2024crag} applies a
post-retrieval corrector.
None of these train a multi-strategy query-type router on a
publicly labeled benchmark.

\paragraph{LLM routing:}
RouteLLM \citep{ong2025routellm} and FrugalGPT
\citep{chen2023frugalgpt} show that lightweight classifiers
can halve API costs when routing between strong and weak models.
RouterBench \citep{hu2024routerbench} demonstrates that KNN and
MLP routers on sentence embeddings are competitive, and Hybrid
LLM \citep{ding2024hybridllm} cuts large-model calls by 40\%
with no quality loss using a DeBERTa router.
We apply the same classifier-based paradigm to RAG
\emph{strategy} selection rather than model selection.

\section{Method}
\label{sec:method}

\subsection{Routing Label and Paradigm Mapping}

We use the query type annotation from RAGRouter-Bench as the
routing target: \texttt{single\_hop} (factual),
\texttt{multi\_hop} (reasoning), and \texttt{summary}
(summarization).
To simulate cost savings, we map each predicted query type to
a recommended paradigm following \citet{jeong2024adaptiverag},
who show that factual questions are well-served by single-step
retrieval, multi-hop questions benefit from richer retrieval,
and summarization queries require iterative retrieval:
\textit{single\_hop}$\to$\textsc{NaiveRAG},
\textit{multi\_hop}$\to$\textsc{HybridRAG},         \textit{summary}$\to$\textsc{IterativeRAG}.

We stress that this mapping is a literature-motivated
simplification for cost estimation only.
The benchmark paper itself shows that optimal paradigm selection
requires both query type and corpus-side signals
\citep{wang2026ragrouterbench}; our classifiers operate on query
text alone, making corpus-side effects a limitation we address
in Section~\ref{sec:limitations}.

\subsection{Feature Sets}

All features are extracted from query text at routing time,
requiring no retrieval or LLM calls.

\paragraph{Lexical (TF-IDF):}
Unigram and bigram TF-IDF with sublinear term-frequency scaling,
a 3,000-feature vocabulary ceiling, and minimum document
frequency of 2.

\paragraph{Semantic (MiniLM Embeddings):}
Each query is encoded with \texttt{all-MiniLM-L6-v2}
\citep{reimers2019sbert}, producing 384-dimensional dense vectors.
This 22M-parameter model runs on CPU with no GPU required.

\paragraph{Structural (Hand-crafted):}
Twenty-three features: query length, character count, average
word length, question-word type (one-hot: who / what / when /
where / why / how / which), presence of negation, approximate
named-entity count, clause count, and binary flags for
comparative, temporal, aggregation, causal, and procedural
patterns.
Question-word type has been shown to correlate with retrieval
complexity by \citet{jeong2024adaptiverag}.

\subsection{Classifiers and Evaluation}

Five classifiers are trained on each feature set:
Logistic Regression (L2, $C$=1.0), SVM (RBF, $\gamma$=scale),
Random Forest (200 trees), KNN ($k$=7, cosine), and
MLP (256--128 units, early stopping).
Dense feature sets are z-score standardized before training.
We report macro-averaged F1 (macro-F1) as the primary metric
to account for label imbalance, with accuracy alongside it.
All results are from 5-fold stratified cross-validation.

\paragraph{Cost simulation:}
We use the paradigm cost ratios reported by
\citet{wang2026ragrouterbench} to compute token savings post-hoc,
without any LLM calls.
Savings are relative to always routing to \textsc{IterativeRAG}:
\[
  \text{Savings} =
  \frac{C_{\textsc{IterativeRAG}} - C_{\text{router}}}
       {C_{\textsc{IterativeRAG}}} \times 100\%,
\]
where $C_{\text{router}} = \sum_{i=1}^{N} c(\hat{y}_i)$
is the sum over all $N$ queries of the cost ratio of the
predicted paradigm $c(\hat{y}_i)$, pooled across all
cross-validation folds. $C_{\textsc{IterativeRAG}} = 3.5N$.

The \emph{perfect-label reference} assumes every query is
routed to its ground-truth type's paradigm:
$C_{\text{ref}} = \sum_{i=1}^{N} c(y_i) = 0.529N(1.4) +
0.171N(2.8) + 0.300N(3.5) = 2.269N$,
giving reference savings of
$(3.5 - 2.269)/3.5 \approx 35.2\%$.
This methodology follows RouteLLM \citep{ong2025routellm} and
RouterBench \citep{hu2024routerbench}.

\section{Results}
\label{sec:results}

\subsection{Routing Performance}

Table~\ref{tab:main} reports routing accuracy and macro-F1 for
all 15 combinations.
TF-IDF with SVM achieves the best overall result
(Acc=93.2\%, F1=0.928), outperforming all structural and
semantic configurations.

\begin{table}[t]
\centering
\small
\setlength{\tabcolsep}{3.5pt}
\caption{Routing accuracy (Acc, \%) and macro-F1 under 5-fold
cross-validation. Best result per metric in \textbf{bold}.
Majority baseline always predicts \textit{single\_hop} (52.9\%).}
\label{tab:main}
\begin{tabular}{lcccccc}
\toprule
 & \multicolumn{2}{c}{\textbf{TF-IDF}} &
   \multicolumn{2}{c}{\textbf{MiniLM}} &
   \multicolumn{2}{c}{\textbf{Structural}} \\
\cmidrule(lr){2-3}\cmidrule(lr){4-5}\cmidrule(lr){6-7}
\textbf{Classifier} & Acc & F1 & Acc & F1 & Acc & F1 \\
\midrule
Logistic Reg.  & 92.1 & 0.918 & 86.8 & 0.864 & 78.1 & 0.763 \\
SVM            & \textbf{93.2} & \textbf{0.928} & \textbf{90.3} & \textbf{0.897} & 79.1 & 0.774 \\
Random Forest  & 91.9 & 0.914 & 80.7 & 0.784 & 77.8 & 0.752 \\
KNN            & 85.4 & 0.855 & 79.3 & 0.775 & 78.3 & 0.762 \\
MLP            & 92.7 & 0.923 & 90.1 & 0.896 & \textbf{80.3} & \textbf{0.788} \\
\midrule
Majority class & \multicolumn{2}{c}{52.9 / 0.231}
               & \multicolumn{2}{c}{---}
               & \multicolumn{2}{c}{---} \\
\bottomrule
\end{tabular}
\end{table}

\subsection{Simulated Cost Savings}

Table~\ref{tab:cost} reports simulated token savings for the
best configuration per feature set, and for reference baselines.
All classifiers trained on text features achieve savings
well above zero, with macro-F1 ranging from 0.788 to 0.928.
The majority-class baseline (60.0\% savings, F1=0.231) and
the perfect-label reference (35.2\%) together define the
cost--accuracy tradeoff landscape: higher savings are achievable
at low accuracy, while high-accuracy routers settle in the
25--30\% range.

\begin{table}[t]
\centering
\small
\caption{Simulated token savings vs.\ always routing to
\textsc{IterativeRAG}. Paradigm cost ratios from
\citet{wang2026ragrouterbench}; type-to-paradigm mapping
from \citet{jeong2024adaptiverag}.
Perfect-label reference assumes correct type prediction for
every query; it is not an upper bound on savings.}
\label{tab:cost}
\begin{tabular}{lcc}
\toprule
\textbf{Configuration} & \textbf{Savings (\%)} & \textbf{Macro-F1} \\
\midrule
TF-IDF + SVM         & 28.1 & 0.928 \\
MiniLM + SVM         & 27.4 & 0.897 \\
Structural + MLP     & 30.2 & 0.788 \\
\midrule
Majority class       & 60.0 & 0.231 \\
Perfect-label ref.   & 35.2 & 1.000 \\
\bottomrule
\end{tabular}
\end{table}

\subsection{Domain Breakdown}

Table~\ref{tab:domain} shows macro-F1 by domain for the best
classifier under each feature set.
Routing is consistently hardest on medical queries and easiest
on legal queries across all three feature regimes.

\begin{table}[t]
\centering
\small
\caption{Domain-level macro-F1, best classifier per feature set.
Wiki=MuSiQue, Lit=QuALITY, Leg=UltraDomain, Med=GraphRAG-Bench.
\textbf{Bold} indicates best score per column.}
\label{tab:domain}
\begin{tabular}{lcccc}
\toprule
\textbf{Feature Set} & \textbf{Wiki} & \textbf{Lit}
                     & \textbf{Leg}  & \textbf{Med} \\
\midrule
TF-IDF (SVM)        & \textbf{0.926} & \textbf{0.951} & \textbf{0.967} & \textbf{0.803} \\
MiniLM (SVM)        & 0.891 & 0.912 & 0.946 & 0.727 \\
Structural (MLP)    & 0.708 & 0.853 & 0.855 & 0.645 \\
\bottomrule
\end{tabular}
\end{table}

\section{Analysis}
\label{sec:analysis}

\paragraph{Lexical features outperform semantic embeddings:}
TF-IDF achieves 0.928 macro-F1, outperforming MiniLM (0.897) by
3.1 points and structural features (0.788) by 14.0 points.
Query type in this benchmark is strongly signaled by surface
keywords: question-word type, domain terminology, and patterns
such as \emph{summarize} or \emph{compare} appear sufficient to
distinguish factual, reasoning, and summarization queries
without semantic encoding.
That MiniLM underperforms TF-IDF likely reflects that dense
embeddings conflate surface-similar but type-different queries
from different domains, while sparse TF-IDF weights retain
vocabulary signals that differ systematically across query types.

\paragraph{The cost-accuracy tradeoff:}
The perfect-label reference achieves 35.2\% savings under the
assumed type-to-paradigm mapping.
High-accuracy classifiers (TF-IDF+SVM at 28.1\%, MiniLM+SVM at
27.4\%) recover 78--80\% of these reference savings while
maintaining strong routing quality.
Structural+MLP achieves 30.2\% savings at a lower F1 of 0.788:
the reduced routing accuracy shifts some multi-hop and summary
queries toward \textsc{NaiveRAG}, which lowers average cost but
degrades answer quality on complex queries.
The majority-class baseline makes this dynamic extreme: always
predicting \textit{single\_hop} and routing every query to
\textsc{NaiveRAG} achieves 60.0\% savings at a macro-F1 of only
0.231---the highest savings of any configuration and the lowest
accuracy.
This illustrates why cost savings and macro-F1 must be reported
jointly: optimising savings in isolation simply collapses routing
to the cheapest paradigm, a finding directly reinforced by
\citet{wang2026ragrouterbench}'s argument against treating routing
as cost minimisation alone.

\paragraph{Domain routing difficulty:}
Routing is hardest on GraphRAG-Bench (medical), with
TF-IDF+SVM reaching only 0.803 macro-F1, compared to 0.967
on UltraDomain (legal).
The medical corpus in RAGRouter-Bench consists of a single long
document, making all query types draw from the same source; surface vocabulary alone may be less discriminative when corpus structure is homogeneous.
Legal queries follow more formulaic patterns across diverse
documents, producing cleaner lexical signals for type prediction.
The 0.164 gap between the easiest and hardest domain underscores
the benchmark's finding that routing is a query-corpus
interaction problem, not a query-only problem
\citep{wang2026ragrouterbench}.

\section{Limitations}
\label{sec:limitations}

All classifiers operate on query text alone.
\citet{wang2026ragrouterbench} show experimentally that
optimal paradigm selection depends on \emph{query-corpus
interactions}: structural corpus properties (connectivity,
density, clustering coefficient) and semantic properties
(intrinsic dimension, hubness, dispersion) jointly shape
which paradigm performs best.
Our query-side baselines deliberately exclude these signals,
making them a floor rather than a ceiling for routing
performance on this benchmark.
The type-to-paradigm mapping used for cost simulation is a
literature-motivated simplification; at the query-corpus
interaction level, the best paradigm for a
\textit{multi\_hop} query may differ substantially across
corpora.
The perfect-label reference does not represent an upper bound
on savings: a routing policy that ignores label fidelity can
achieve higher savings (as the majority-class baseline
demonstrates at 60.0\%) at the expense of answer quality.
Finally, all models are evaluated within RAGRouter-Bench's
four domains; out-of-distribution generalisation is untested.

\section{Conclusion}
\label{sec:conclusion}

We present the first query-type routing classifier baselines
on RAGRouter-Bench, evaluating 15 classifier--feature
combinations for paradigm selection.
TF-IDF with an SVM achieves 0.928 macro-F1 and 93.2\%
accuracy, simulating 28.1\% token savings versus always using
the most expensive paradigm---79\% of the 35.2\% savings
achievable under perfect type-faithful routing.
Lexical features outperform sentence embeddings by 3.1
macro-F1 points, showing that surface keyword patterns are
strong routing signals for this benchmark.
A joint analysis of cost and accuracy reveals that optimising
savings alone is misleading: the majority-class baseline achieves
60.0\% token savings at a macro-F1 of only 0.231 by routing
every query to the cheapest paradigm regardless of complexity.
These baselines quantify the routing signal available from
query text alone, and the gap to the perfect-label reference
highlights the value of incorporating corpus-side signals in
future work.
\bibliography{refs_routing}

\end{document}